\documentstyle[11pt,flushrt,aaspp4]{article}
\def\et{et al.\ }
\def\xte{{\it RXTE}}
\def\exosat{{\it EXOSAT}}

\def\Msun{\hbox{$\rm\thinspace M_{\odot}$}}
\begin{document}
\title{A Cutoff in the X-ray Fluctuation Power Density Spectrum of 
	the Seyfert 1 Galaxy NGC 3516}
\author{Rick Edelson}
\affil{ Astronomy Department; University of California;
        Los Angeles, CA 90095-1562; USA}
\affil{X-Ray Astronomy Group; Leicester University; Leicester LE1 7RH;
        United Kingdom}
\affil{Eureka Scientific; 2552 Delmar Ave.; Oakland, CA 94602-3017; USA}
\authoremail{rae@star.le.ac.uk}
\and
\author{Kirpal Nandra}
\affil{NASA/Goddard Space Flight Center; Laboratory for High Energy
        Astrophysics; Code 662; Greenbelt, MD 20771; USA}
\altaffiltext{1}{Universities Space Research Association}
\authoremail{nandra@phoenix2.gsfc.nasa.gov}

\begin{abstract}
During 1997 March--July, \xte\ observed the bright, strongly variable
Seyfert 1 galaxy NGC 3516 once every $\sim$12.8~hr for 4.5~months and 
nearly continuously (with interruptions due to SAA passage but not Earth 
occultation) for a 4.2~day period in the middle.  
These were followed by ongoing monitoring once every $\sim$4.3~days.  
These data are used to construct the first well-determined X-ray 
fluctuation power density spectrum (PDS) of an active galaxy to span more 
than 4 decades of usable temporal frequency.  
The PDS shows no signs of any strict or quasi-periodicity, but does show a 
progressive flattening of the power-law slope from --1.74 at short time 
scales to --0.73 at longer time scales.  
This is the clearest observation to date of the long-predicted cutoff in the 
PDS.  
The characteristic variability time scale corresponding to this cutoff 
temporal frequency is $\sim$1~month.  
Although it is unclear how this time scale may be interpreted in terms of a 
physical size or process, there are several promising candidate models.  
The PDS appears similar to those seen for Galactic black hole candidates 
such as Cyg~X-1, suggesting that these two classes of objects with very 
different luminosities and putative black hole masses (differing by more 
than a factor of $10^5$) may have similar X-ray generation processes and 
structures.
\end{abstract}


\section{ Introduction }

X-ray variability can, in principle, provide strong constraints on the
physical conditions in the centers of Active Galactic Nuclei (AGN).
Early observations of rapid variability in Seyfert galaxies implied large
luminosity densities, providing the first compelling argument for the
standard black hole/accretion disk model (Rees 1984) and against
starburst models (Terlevich \et 1995).  In some models, the X-ray
source is the primary emission component, so X-ray variability may be
the most direct way to probe the smallest accessible regions of the
central engine.

Measurement of a ``characteristic variability time scale" may allow
determination of the source size, luminosity density, and black hole
mass.  Early attempts to do this with unevenly sampled data involved
measuring the ``e-folding time" (or the similarly defined doubling
time), defined as the minimum observed value of $ t_{1/e} = |[d {\rm
ln}F/dt]^{-1}| $.  However, this quantity is unstable, depending
strongly on sampling length and signal-to-noise ratio (Press 1978).
There have also been claims for quasi-periodic oscillations in AGN
(e.g., Papadakis \& Lawrence 1993, 1995, Iwasawa \et 1998), but these 
have been disputed (e.g., Tagliaferri \et 1996), and there is certainly 
no consensus that such behavior has been observed in any AGN.

The best way to measure the characteristic variability time scale, and 
to characterize AGN variability in general, is to determine the
fluctuation power density spectrum (PDS).  Because of its highly
eccentric orbit, \exosat\ provided the most useful AGN light curves
gathered until recently with its uninterrupted 2--3 day ``long-looks."
The \exosat\ long-look PDS had a usable dynamic range of about 1.7 decades 
of temporal frequency ($ \sim 10^{-3} $ to $ 2 \times
10^{-5} $~Hz, or $ \sim 10^{3}$ to $ 5 \times 10^4 $ sec).  
The upper frequency
limit is the smaller of the Nyquist frequency (half the inverse of the
fundamental sampling period) and the frequency at which the variations
are typically greater than the noise, and the lower limit corresponds
to $\sim$1/3 of the observation length (e.g., Press 1978).  These PDS
generally rise smoothly to longer time scales as a power-law ($ P(f)
\propto f^a $, where $P(f)$ is the power at temporal frequency $f$),
with no features or signs of periodicity.  All were apparently
consistent with a single form, so-called ``red noise" because the high
temporal frequency slopes were steep ($ a \approx -1.5 $; Green \et
1993; Lawrence \& Papadakis 1993).  A turnover (which would be
identified as a characteristic variability time scale) must occur at
some longer time scale or the total variability power would diverge,
but these PDS were not able to detect it because of the lack of low
temporal frequency coverage.  Attempts have been made to constrain
this cutoff by combining data from a number of unrelated observations
(e.g., McHardy 1989).  The most convincing application to date was for
NGC 4151 (Papadakis \& McHardy 1995), which found evidence that the PDS 
was cut off at a time scale of $\sim 2$ weeks.  These previous
efforts utilized highly unevenly sampled data, however, and therefore
required complex and non-standard techniques that were not able
to directly determine the PDS, leaving the results with considerable 
uncertainty.  
Here we present the results of more standard PDS analysis 
of a new body of near-evenly sampled data.

\section{ Observations }

It was only after \exosat's mission had ended that the importance of
evenly-sampled data and simultaneously probing long and short time 
scales became apparent.  All
X-ray satellites launched since then have had low-Earth orbits, for
which Earth occultation generally corrupted short-term light
curves beyond recovery.  However, even though it too had a low-Earth
orbit, the launch of the {\it Rossi X-ray Timing Explorer} (\xte)
opened up interesting, new possibilities.  It's high throughput, fast
slewing and flexible scheduling make it ideal for obtaining even sampling 
on long time scales.  Most importantly for this project, it has a large
``continuous viewing zone," defined as a region of sky near the
orbital pole for which a source is visible over a number of
consecutive orbits without interruption due to Earth occultation.
Furthermore, any source in such a region is also observable for
at least some part of each orbit, at all times of the year.  
This makes it possible to evenly sample both long and short time scales
with a single instrument.

\subsection{ NGC 3516 }

We searched this region of sky and found, by fortunate coincidence,
that it contained NGC~3516.  
This bright ($B \approx 13 $) object was one of the original
Seyfert galaxies (Seyfert 1943).  It is bright in the X-rays as well,
with $ F({\rm 2-10~keV}) \approx 1.3-8 \times 10^{-11} $~erg
cm$^{-2}$~sec $^{-1}$ (Ghosh \& Soundararjaperumal 1991; Kolman \et
1993; Nandra \et 1997a).  
It shows a complex spectrum, and a simple power-law plus cold absorption 
model is inadequate to describe it.
There is strong evidence for further spectral complexity in this source, 
including a ``warm absorber'' (Kolman \et 1993; Kriss \et 1993; George \et 
1998), a broad, variable iron K$\alpha$ line (Nandra \et 1997a,b) and 
probably also Compton reflection (Kolman \et 1993; Nandra \& Pounds 1994).  
The X-ray continuum flux of NGC 3516 is strongly variable, as is its spectrum 
(Kolman \et 1993; Kriss \et 1993).  
The ultraviolet continuum flux and C~IV absorption also show variations on 
time scales as short as a few days to a week (Voit, Shull \& Begelman 1987; 
Koratkar \et 1997).
Because it is so bright an strongly variable, and most importantly because of 
its favorable location in the sky, NGC~3516 was chosen for this experiment.

\subsection{ Sampling Pattern }

The sampling pattern was optimized to take full advantage of the unique
attributes of \xte\ to obtain variability information spanning time scales 
from minutes to months.
NGC~3516 was observed once every $\sim$12.8~hr (=8 orbits) for 124 epochs, 
then quasi-continuously for 4.2~days, then again once every $\sim$12.8~hr 
for another 124 epochs.
(Combining the periodic sampling with snippets from the continuous period
yielded 256 nearly-evenly-sampled scheduled epochs.)
These were followed immediately by observations once every $\sim$4.3~days 
(=64 orbits) that are planned to continue for the lifetime of the 
satellite.

The goal was for the sampling to be as close as possible to strictly
periodic on each of three progressively shorter and more densely
sampled time scales.  
This experiment came much closer than any previous effort, although certain 
unavoidable perturbations must be noted.  
During the continuous sampling, the light curve was interrupted by SAA 
passage, and a few time-critical observations of other sources also caused 
short interruptions.  
The total time lost to all of these effects was 69~ksec (19\% of the total 
duration).  
The 12.8~hr sampling was interrupted three times for a total of 10 epochs
(3.9\%), the longest being 6 epochs when the satellite went into safe
mode from May 31 06:24 to June 2 22:24.  
For the longest sampling, there were a total of 3 interruptions (2.3\%).  
The real sampling pattern was matched as closely as possible to the 
ephemeris.
The mean deviation between the ephemeris and observation times was
0.36~hr (2.8\% of a fundamental cycle) for the medium time scale and
0.98~hr (1.0\%) for the long time scale sampling.  
(The short time scale data were by definition tied to the ephemeris with no 
deviation.)

The sampling parameters are summarized in Table~1.  For each time
scale (short, medium and long) for which there is a near-evenly sampled
light curve, column~2 gives the range of observing dates, column~3,
the mean sampling interval, column~4, the number of points, and
column~5, the usable temporal frequency range, respectively.

\begin{deluxetable}{ccccc}
\tablewidth{0pc}
\tablenum{1}
\tablecaption{Sampling Parameters \label{tab1}}
\small
\tablehead{
\colhead{Time} & \colhead{} & \colhead{Sampling} &
\colhead{Number} & \colhead{Temporal Frequency} \\
\colhead{Scale} & \colhead{Range of Dates} &
\colhead{Interval} & \colhead{of Points} & \colhead{Range (Hz)} }
\startdata
 Short  & 1997 May 22 00:14 -- 1997 May 26 05:37 & 11.8 min & 512 &
        $ 5.7 \times 10^{-6} - 7.0 \times 10^{-4} $  \nl
 Medium & 1997 Mar 16 00:01 -- 1997 Jul 30 03:46 & 12.8 hr &  256 &
        $ 1.7 \times 10^{-7} - 1.1 \times 10^{-5} $  \nl
 Long   & 1997 Mar 16 00:01 -- 1998 Sep 12 16:06 & 4.27 days &  128 &
        $ 4.2 \times 10^{-8} - 1.3 \times 10^{-6} $  \nl
\enddata
\end{deluxetable}

\section{ Data }

\subsection{ Data Reduction }

The \xte\ Proportional Counter Array (PCA) consists of 5 collimated
Proportional Counter Units (PCUs), sensitive to X-rays in a nominal
2-100~keV bandpass.
Some further details of the PCU and its performance are given by Jahoda
\et (1996).
Here only the PCA {\tt STANDARD-2} data are considered, which have a 
minimum of 16 sec time resolution and full detector and layer 
identification.
The first of the PCU layers is most sensitive to cosmic X-rays and 
therefore only those data are employed herein.
PCU units 3 and 4 were occasionally turned off due to performance problems
and here only data from three of the PCUs (0, 1 and 2) are used.
The analysis is further restricted to the 2-10 keV band, where the PCA is
most sensitive and the systematic errors are best quantified.

The data reduction methods employed here are similar to those described by
Nandra \et (1998).
In particular, binned light curves were initially extracted with 16 sec 
time resolution from the {\tt STANDARD-2} data.
Poor quality data were excluded using the criteria that:
the Earth elevation angle was greater than 5 degree; the offset between the
pointing position and the optical position of NGC 3516 was less than 0.01
degree and to exclude periods where the anti-coincidence rate in the
propane layer of the PCUs was abnormal.
This last criterion, used to identify anomalous electron flares, is more
fully described by Nandra \et (1998).
The relatively lax Earth elevation criterion was used to maximize the 
temporal coverage without significantly degrading the data quality.
The 16~sec sampled data were then rebinned to longer sampling intervals as
discussed below.
The light curve is shown in Figure~1.

\placefigure{1}

\subsubsection{ Background Model }

Background estimation represents the largest uncertainty for
relatively weak sources such as NGC 3516.  
The three main components of the layer 1 background are those related to 
diffuse (sky) X-rays, a term related to many of the particle 
anti-coincidence rates and another that arises from decays induced by the 
South Atlantic Anomaly (SAA). 
Two background models, termed ``{\tt SKY\_VLE}" and ``{\tt L7},'' were 
considered. 
Only data using the latter model are used for the variability analysis, as 
the PCA team consider it more reliable, a conclusion that is confirmed below. 
The results obtained with each model are, however, sufficiently similar 
that they do not affect the conclusions presented in this paper.
Both models are based on analysis of ``sky'' data from a number of
pointings toward regions that are believed to be free of bright 
X-ray sources.

The``{\tt SKY\_VLE}" model estimates the particle background by
relating it to the Very Large Event (VLE) rate of the PCA.  
About half of \xte's orbits pass through the SAA each day, inducing 
background terms that decay exponentially.  
The SAA-related component was estimated by splitting the day into two 
halves.  
Orbits in the non-SAA half were assumed to be unaffected by previous 
passages, which is only valid if there are no long-term decays in the 
background.  
Under this assumption, background data were then used to estimate a 
component related to the VLE rate.  
The VLE model was then applied to the SAA orbits and the residual flux 
assumed to be that induced in the SAA.  
This flux was then correlated with satellite orbital parameters 
({\tt BKGD\_PHI} and {\tt BKGD\_THETA}) to define the model.  
Then, the background for the current observations is estimated by taking the 
assumed spectral forms of the VLE and SAA-related components and combining 
them according to the observed VLE rates and positional parameters.

The ``{\tt L7}'' model employs other PCA event rates that are directly
related to both the particle and activation rates by a scaling factor.
This leaves a residual induced term in the background that is modeled
as an exponential decay with an e-folding time of 240 min. 
The amplitude of this term is determined by integrating the rates from
particle monitors on the High Energy X-ray Timing Experiment (HEXTE) on 
board \xte, which are used to estimate the ``dose'' in the SAA.

We became aware of the importance of background subtraction during the
campaign, and therefore began to accumulate offset data after (and
sometimes before) the snapshots of NGC 3516 in a region of sky close
by, but believed free of bright sources.  
These offset data were used to assess the systematic errors in the 
background subtraction methods.
The {\tt L7} model gave the best subtraction of the offset data, and the 
background-subtracted offset data for that model is shown at the bottom of 
the top panel in Figure~1. 
Even for this best model, the mean is non-zero and the variance is larger 
than would be expected from statistical fluctuations. 
The mean and excess deviation are 0.87 and 0.39~c/s for the {\tt L7} model 
and 0.45 and 0.78~c/s for the {\tt SKY\_VLE} model. 
The lower excess variance shows that that the {\tt L7} is significantly 
better and that is why it was used in all subsequent analyses.

The corresponding section of the NGC~3516 light curve has an RMS
deviation of 4.4~c/s.  
The ratio of the squares (variances) is $\sim$125, meaning that (in the 
worse case) typically less than 1\% of the observed variability power is due 
to errors in the background model. 
The raw offset data have been corrected with the background-subtracted 
offset data, to see if there are any residual trends that have been poorly 
subtracted. 
The data showed no significant correlation, further strengthening 
confidence in the background subtraction. 
We therefore interpret the data under the assumption that the observed 
variability is dominated by intrinsic changes in the flux of NGC~3516, but 
caution that systematic errors in the background model could potentially 
cause subtle, low-level problems.

\subsection{ Data Analysis }

The measurement of the PDS took two steps: 
First, the short, medium, and long time scale data were each separately 
made suitable for analysis and the individual PDS were measured, as 
discussed in the next subsection.  
Then, the separate PDS were combined to produce a single PDS, as discussed in 
the second subsection.

\subsubsection{Construction of the Individual PDS}

The uneven sampling that made it so difficult for previous campaigns to 
estimate the PDS was much less of a problem here.
The data were linearly interpolated (without adding Poisson noise) across 
the relatively rare data gaps, which generally spanned less than a few 
contiguous epochs because they tended to be short and spread quasi-randomly 
throughout the observation.
This produced three light curves: (1) with 512 points sampled every 
11.8~min, (2) with 256 points sampled once every 12.8~hr, and (3) with 128 
points sampled once every 4.3~days.
Because the departures from the ephemeris were almost always small
(typically only a few percent), interpolating and rebinning the data 
to an even grid would have made little difference (less than 0.3\% on 
average, which is much less than the statistical or systematic errors).
Thus, the ephemeris times and not the actual times were used in the 
analysis, with no interpolation.

PDS were then derived using standard methods from Numerical Recipes
(Press \et 1992) and the logarithms of both frequency and power
were taken.
The zero-power and next two lowest-frequency points of each PDS were 
ignored in the further analyses because they tend to be extremely noisy 
(see Press 1978 for details).
The remaining points were binned every factor of two (0.3 in the logarithm), 
to reduce the noise and allow estimation of error bars.
That means that the first binned point was derived by binning the two
remaining lowest frequency points, the second by binning the next four
points, etc.

The PDS were calculated using several window functions, $w_j$, to test for
the effect of ``red-noise leak'' (e.g., Deeter \& Boynton 1982).
These included the square window function
\begin{equation}
w_j=1,
\end{equation}
the Welch window function,
\begin{equation}
w_j= 1 - \left(\frac{j-\frac{1}{2} N}{\frac{1}{2} N}\right)^{2},
\end{equation}
the Hann window function, 
\begin{equation}
w_j= \frac{1}{2} \left[ 1 - {\rm cos}\left(\frac{2\pi j}{N}\right)\right],
\end{equation}
and the Bartlett window function,
\begin{equation}
w_j= 1 - \left|\frac{j-\frac{1}{2} N}{\frac{1}{2} N}\right|.
\end{equation}
The individual PDS slopes differed only slightly depending on the window
function. 
Results are presented in Figure~2 for the Bartlett window. 

\placefigure{2}

Power-law models were fitted to each PDS to estimate the slope, as
shown in Table~2: column~2 gives the mean deviation of the data from
the ephemeris, column~3, the amount of data lost, column~4, the fitted
slope (and uncertainty) of the power-law fit to each data set.  
The slope and uncertainty were measured from an unweighted, least-squares
fit to the logarithmically binned PDS.
The fractional variability ($F_{var}$) is given in the last column.  
Unlike Edelson \et (1996), this number is {\it not} corrected for 
intrinsic errors, because the uncertainties are dominated by systematic 
errors that have a different (and unknown) effect on different time scales, 
and in any event, the correction for these effects is small compared to the 
total observed value.

\begin{deluxetable}{ccccc}
\tablewidth{0pc}
\tablenum{2}
\tablecaption{Variability Parameters \label{tab2}}
\tablehead{
\colhead{Time} & \colhead{Mean Deviation} & \colhead{} &
\colhead{Power-Law} & \colhead{Fractional} \\
\colhead{Scale} & \colhead{from Ephemeris} & \colhead{Data Lost} &
\colhead{Slope ($a$)} & \colhead{Variability ($F_{var}$)} }
\startdata
Short  & ---             & 69 ksec  (19.1\%) & $-1.74 \pm 0.12 $ &  7.9\% 
\nl
Medium & 0.36 hr (2.8\%) & 10 points (3.9\%) & $-1.03 \pm 0.06 $ & 30.1\% 
\nl
Long   & 0.98 hr (1.0\%) &  2 points (2.3\%) & $-0.73 \pm 0.12 $ & 31.6\% 
\nl
\enddata
\end{deluxetable}

\subsubsection{Construction of the Combined PDS}

\begin{deluxetable}{lcc}
\tablewidth{0pc}
\tablenum{3}
\tablecaption{PDS Fit Parameters \label{tab3}}
\tablehead{
\colhead{Description} & \colhead{Parameter} & \colhead{Value} }
\startdata
High frequency slope	   & $a$	   & $-1.76$ \nl
Cutoff frequency	   & $f_c$	   & $ 4.14 \times 10^{-7}$ Hz \nl
Cutoff time scale	   & $t_c(=1/f_c)$ & 27 days \nl
Normalization coefficients & $C_1$	   & 8.54 \nl
			   & $C_2$	   & 354 \nl
			   & $C_3$	   & 646 \nl
\enddata
\end{deluxetable}

The short, medium, and long time scale PDS were then combined to form
a single PDS spanning four decades of usable temporal frequency.  
As shown in Table~2, the PDS shows a highly significant (8$\sigma$) 
systematic flattening from $ a = -1.74 \pm 0.12$ to $-0.73 \pm 0.12$ as 
one goes from short to long time scales.  
These values were derived using the Bartlett window, but a similar 
difference of ($\Delta a \approx 1$) was also seen for the other window 
functions.
Because this same difference is seen for all window functions, we conclude 
that red-noise leak cannot be responsible for the change in PDS slope.

Thus, a more complex model than a single power-law was required to describe 
the PDS.
All three PDS were simultaneously fitted with a model of a power-law 
that dominates at high frequencies, but cuts off to a slope of $ a = 0 $ at 
very low temporal frequencies.
We used the function 
\begin{equation}
 P(f) = C_1 / ( 1 + f / f_c )^a,
\end{equation}
where $P(f)$ is the fluctuation power at temporal frequency $f$, $a$ is the
power-law slope at high temporal frequencies, $f_c$ is the ``cutoff
frequency," well below which the PDS flattens to a slope of zero, and $C_1$
is the normalization. 
Arbitrary, free {\it relative} normalizations ($C_2$, $C_3$) were allowed 
between the three individual PDS. 
Including these two scaling factors, there are a total of five free 
parameters in the fit.
The fit was performed by minimizing the Whittle negative log likelihood
function with the result  shown in Figure~3 and Table~3.
The high-frequency slope is $ a = -1.76$, similar to the slope derived 
using the high-frequency PDS alone. 
The cutoff frequency $f_{c}$ corresponds to a time scale of 27~days.

Although it is clear that the PDS is not consistent with a single power-law, 
the exact shape of the turnover is not well-constrained and the low-frequency 
behavior is undetermined.
The fit assumed $ a \rightarrow 0 $ for $ f \ll f_c $, but in fact it could
be any slope with $ a > -1 $.
(For $ a \le -1 $, the total variability power would diverge.)

It is highly unlikely that the background problems discussed earlier could 
be responsible for the flattening in the PDS, because, as mentioned 
earlier, background errors contribute less than 1\% of the total variability 
power.
Furthermore, examination of Figure~1 shows that the background errors
are correlated with each other; that is, their power is greater on long 
time scales than on short time scales.  
If anything, this would add more power at long time scales where the 
excursions are largest and relatively less at short time scales.  
This would {\it steepen} the PDS at long time scales, but this is the 
opposite of the {\it flattening} that is actually observed.

Finally, the $F_{var}$ values in Table~2 give independent 
evidence for a cutoff in the PDS at the longest time scales sampled.  
$F_{var}$ rises from 7.9\% to 30.1\% for short and medium time scale data 
taken over a period of 4.2~days and 4.5~months, respectively.  
However, it shows little further rise in the long time scale data, 
flattening out to 31.6\% for sampling over 1.5~years.
Again, visual examination of Figure~1 suggests that there is very
little extra variability power on the longest time scales sampled.

\section{ Discussion }

We have obtained near-evenly sampled light curves of NGC 3516 over an
unprecedented range of time scales. 
We have used these data to construct a PDS over 4 decades of temporal 
frequency. 
No sharp features were observed in the PDS. 
On short time scales, NGC 3516 exhibits ``red-noise.''
The high-frequency slope is consistent with the mean values from 
{\it EXOSAT} of $a = -1.7 $ and $a=-1.55 $ derived by Green \et (1993) and 
Lawrence \& Papadakis (1993), respectively.
The extension of the PDS to lower frequencies afforded by the medium
and long-time scale data has shown new results, however. 
The PDS is seen to progressively flatten at longer time scales and we 
interpret the overall PDS as possessing a characteristic variability time 
scale corresponding to the frequency of the turnover. 
Determining the variability time scale from these data is difficult, but 
our best estimate is $t_{c} \approx 1 $~month.

\subsection{ Physical Time Scales }

Here we state some characteristic time scales within the source, which
may possibly be related to the observed flattening in the PDS.  
The fastest possible variability time scale for a coherent, 
isotropically-emitting region is the light-crossing time:
\begin{equation}
t_{\rm lc} = 0.011 \ (M_{BH}/10^7 M_\odot) \ (R/10 R_S) \ \rm day.
\end{equation}
Here $M_{BH}$ is the black hole mass, $R$ is the distance from the
center of mass and $R_{S}$ is the Schwarzschild radius.
In a Comptonizing cloud this can be modified by scattering effects.
The effect is negligible for low optical depth ($ \tau \ll 1 $), but for
$ \tau > 1 $ the time scale is increased by approximately $ \tau^{2} $. 

From Kepler's Third Law, the matter orbital time scale is:
\begin{equation}
t_{\rm orb} = 0.33 \ (M_{BH}/10^7 M_\odot) \ (R/10 R_S)^{3/2} \ \rm
day.
\end{equation}

There are also several time scales related to the accretion disk.
Specifically, for an $\alpha$-disk (Shakura \& Sunyaev 1973), relevant
time scales are (Maraschi, Molendi \& Stella 1992; Treves, Maraschi \&
Abramowicz 1988):
\begin{itemize}
\item{Thermal:}
\begin{equation}
t_{\rm th} = 5.3 \ (\alpha/0.01)^{-1} \ (M_{BH}/10^7 M_\odot) \
(R/10 R_S)^{3/2} \ \rm day,
\end{equation}
\item{Sound Crossing:}
\begin{equation}
t_{\rm sound} = 33 \ (R/100 H) \ (M_{BH}/10^7 M_\odot) \ (R/10 R_S)^{3/2} \ 
\rm day, 
\end{equation}
\item{Radial Drift/Viscous ($\dot{M}$):}
\begin{equation}
t_{\rm drift} = 53000 \ (R/H)^{2} \ (\alpha/0.01)^{-1} \
(M_{BH}/10^7 M_\odot) \ (R/10 R_S)^{3/2} \ \rm day.
\end{equation}
\end{itemize}
Here, $\alpha$ is the viscosity parameter of the disk and $H$ is the
scale height of the disk.  

All of the time scales above depend on the mass of the black hole and the 
Distance from the center, in Schwarzschild radii, at which the X-rays are 
emitted. 
Black hole masses in local AGN have been estimated by various techniques 
and generally lie in the range $10^{6-9}$~\Msun\ (Ford \et 1994; Miyoshi \et 
1995; Magorrian \et 1998).  
The likely mass for NGC 3516 is in the range $10^{{7-8}}$~\Msun, 
considering its luminosity. 
This implies that $t_{\rm lc}$ is much shorter than the turnover time scale 
$t_{\rm c}$.
An even more compelling argument against the light-crossing time is that 
reverberation mapping indicates that the broad, optical-line region (BLR) 
in NGC 3516 is light-days to light-weeks across (Wanders \et 1993). 
It is difficult envisage a scenario whereby the BLR is smaller than the 
continuum region that excited it. 
By similar arguments, the presence of a broad, iron K$\alpha$ line in NGC 
3516 (Nandra \et 1997a,b) indicates that the X-ray continuum-producing 
region lies inside the 100 $R_{{\rm S}}$ within which the bulk of that 
emission line is produced. 
We therefore conclude that the turnover cannot be due purely to
light-travel time effects in the X-ray source.

\subsection{Model Implications}

The observed cutoff should be related to the fundamental physics that
generates the variability.  The process by which X-rays are produced 
in AGN is still not known, but typically models have concentrated on
Compton upscattering of ultraviolet ``seed" photons that probably
arise in an accretion disk (e.g., Haardt \& Maraschi 1991, 1993;
Zdziarski \et 1994; Stern \et 1995).

In such a Comptonizing plasma, which is posited to be the X-ray
producing region in AGN, optical depth effects can smear out the
variability, changing the power spectrum. 
For example, if the PDS was originally white noise, passage through such a 
plasma would introduce a cutoff, turning the high frequency PDS into ``red 
noise'' (e.g. Brainerd \& Lamb 1987; Kylafis \& Klimis 1987;
Bussard \et 1988; Lawrence \& Papadakis 1993). 
In such models, the turnover would be identified with the size of the 
scattering region, modified by optical depth effects. 
Identification of the observed turnover with such a process would require 
both a large region and a large optical depth.  
Even at the limits of plausible parameter space ($M_{BH} \approx 
10^{{8}}$~\Msun\ and $R \approx 100 R_{\rm S}$, however, an optical depth 
of order $\tau \sim 10$ would be required to match the observed value of 
$t_{\rm c}$. 
Current models of the X-ray spectra of AGN usually assume lower optical 
depths, and in fact most are optically thin (Haardt \& Maraschi 1991, 1993; 
Stern \et 1995; Zdziarski \et 1995). 
Some AGN emission models have optical depths as large as a few (e.g., 
Titarchuk \& Mastichidias 1994).

It is perhaps more likely, however, that the turnover corresponds to
some different time scale. 
A specific model that has been proposed to explain the variability, the PDS 
and the variation of the PDS with luminosity is the so-called ``bright 
spot'' model of Bao and Abramowicz (1996). 
In this model, active regions on the surface of a rotating accretion disk 
produce the observed variability. 
In such a model the relevant turnover time scale could perhaps be 
identified with the $t_{\rm orb}$, marginally consistent with the observed 
cutoff for the extremes of parameter space mentioned above if the emission is 
produced very far out in the disk.

Mineshige \et (1994a,b) have presented a model in which the central
regions of the accretion disk are in a ``self-organized critical''
(SOC) state. Their predicted PDS are qualitatively similar to those
observed here, with PDS slopes of $ a \approx 1.6 $ at high frequencies 
and a turnover at longer time scales. 
In their model the central accretion disk exists in a near-critical state; 
when the mass density exceeds the critical value an ``avalanche'' of 
accretion occurs, emitting X-rays. 
It is these patchy, unstable regions that produce the X-rays, as well as 
the high frequency noise. 
The turnover time scale corresponds to $t_{\rm drift}$ for the largest blobs. 
In a smoothly-accreting system, such as Mineshige \et assume exists beyond 
the SOC region, $t_{\rm drift}$ is clearly too long to be associated with 
our turnover time scale. 
However, in the critical region in which the putative discrete blobs are 
present, the radial drift time scale can be much shorter, by a factor 
$\sim H/R$. 
Depending on the viscosity parameter, the accretion time scale can then 
approach the sound speed and match the observed cutoff.

Neither the ``bright spot'' nor the SOC models are clear about the
acceleration mechanism or instability that causes the active regions
to form, and it is quite plausible that the turnover is connected to
these more fundamental mechanisms.  
Energy deposition into coronal hotspots might arise from thermal or 
acoustic instabilities in the disk whereby the turnover may be identified 
with $t_{\rm th}$ or $t_{\rm sound}$, which are of the correct order.
At this stage, it is most important that specific physical models of 
variability -- which have been scarce to date -- be constructed and PDS 
predicted for them. 
These can then be compared explicitly with these new data.

\subsection{ Comparison with Galactic X-Ray Binaries }

The PDS of NGC 3516 looks remarkably like that of Cyg X-1 and other
Galactic X-ray binaries (XRBs) in the ``low'' or hard state (see,
e.g., Belloni \& Hasinger 1990; Miyamoto \et 1992; van der Klis
1995). 
These sources exhibit a ``red noise'' spectrum at high frequencies with 
slopes $a$ between $-1$ and $-2$. 
As observed in NGC 3516, they flatten to lower
frequencies, with $ a \approx 0 $ and cutoff frequencies of order 0.1~Hz
(XRBs can also exhibit a very low-frequency noise component, which shall be 
ignored for the purposes of this discussion.)
The analogy between these sources and NGC 3516 is clear.

A simple prediction would be that the cutoff frequency would scale
with physical size and therefore black hole mass, as one would expect,
for example, for the light crossing or orbital time scales. 
In the absence of other specific models, one can relate the ratios of the 
black hole masses in Cyg X-1 and NGC 3516 by a simple scaling law.  
The cutoff frequency for Cyg X-1 varies between $f_{\rm c} = 0.04-0.4$~Hz 
(Belloni \& Hasinger 1990) while for NGC~3516, it appears to be of order 
$ f_{\rm c} \approx 4 \times 10^{-7} $~Hz.  
The ratio is $ 10^5 - 10^6 $.  
The mass of the compact object in Cyg X-1 is thought to be of order
$\sim 10~M_\odot$ by independent arguments (Herrero \et 1995).
Scaling the variability time scale as the mass ($ M \propto R_S
\propto t_c $) yields a mass of $\sim 10^6 - 10^7~M_\odot$ for NGC 3516.
This is to be compared with the estimates above for the black hole
mass, i.e. similar to but perhaps a bit smaller than anticipated.

Of course, scalings and transformations of this kind are not necessarily 
valid unless one knows with which time scale the cutoff is identified.  
If, for example, the relevant quantity is one of the time scales in the
accretion disk, several other parameters come into play which might be
very different when comparing stellar and super-massive black holes.
The luminosities of NGC~3516 and Cyg~X-1 also scale as their cutoff 
frequencies, however, implying a similar Eddington ratio in the two cases.

The similarities in the PDS reinforce the idea that similar physical 
processes may produce the X-rays in AGN and XRBs.  
The hard power-law in XRBs are thought to arise from Comptonization by hot 
electrons and of course similar models have been proposed for AGN. 
(See Van Paradijs 1998 for an excellent review of XRBs.) 
XRB models may deserve some consideration for AGN too. 
For example, Kazanas, Hua \& Titarchuk (1997) have proposed a model for 
XRBs in general and Cyg X-1 in particular, in which the general 
characteristics of the PDS can be produced by photon diffusion through a 
relatively large Comptonizing cloud, with variable density profile.  
As stated above, such Comptonization models require both large source 
regions and high optical depths.  
A prediction of this model is that there should be phase-dependent time 
lags between various energy bands in the X-rays (Hua, Kazanas \& Titarchuk 
1997).
This could be tested by searching the \xte\ data for lags between the hard 
and soft bands, although that will probably have to wait for further 
improvements in the background model and response matrix. 

Along with the broad-band aperiodic variability, XRBs show
quasi-periodic oscillations (QPO), pulsations and bursts. These latter
properties are more easily related to the parameters of the central
source (mass, magnetic field strength, etc.), but unfortunately, their
existence in AGN remains controversial. 
Another interesting property, described for Cyg X-1 above, is that the 
cutoff frequency in XRBs is not constant. 
Prospects for measuring changes in $f_{\rm c}$ in AGN such as NGC 3516 are 
somewhat grim, given that it takes several years just to accumulate 
sufficient data to detect it. 

\section{ Conclusions }

By taking even sampling of the light curves of NGC~3516 on time scales
that ranged from minutes to months, it has been possible to measure the PDS 
from $ 4 \times 10^{-8} $ to $ 7 \times 10^{-4} $~Hz.  
This is more than twice the logarithmic dynamic range of the \exosat\ 
long-looks, which covered  $\le 2 $ decades of usable temporal frequency 
near the high frequency end of this experiment.

The PDS of NGC~3516 is not consistent with a single power-law.
The PDS slope changes from $ a = -1.74 $ at high temporal frequencies
($ 5 \times 10^{-6} - 7 \times 10^{-4} $~Hz) to $ a = -0.73 $ at low
temporal frequencies ($ 4 \times 10^{-8} - 1.3 \times 10^{-6} $~Hz).
Instead, the data are well-fitted by a high-frequency power-law slope of
$ a = -1.76 $, with a cutoff at $ f_c \approx 4 \times 10^{-7} $~Hz.
The PDS slope was assumed to go to a slope of zero for $ f \ll f_c $.
There is no evidence for strict or quasi-periodicity in the PDS.

The observed cutoff time scale is much too long to be associated with
the light-crossing time for any reasonable range of source parameters.  
It is not terribly consistent with the rotating disk model of Bao \& 
Abramowicz (1996) unless the X-rays are produced in the very outer regions
of the disk.  
It could be reconciled with Comptonization in a corona around an accretion 
disk (e.g., Haardt \& Maraschi 1993) if there are substantial optical 
depths effects (Kazanas \et 1997). 
Another suggestion is that it represents the accretion time scale for blobs 
in a self-organized critical disk (Mineshige \et 1994). 
In any event, measurement of the broad PDS and cutoff are basically new
observational results, and at this stage, most models have not been
developed to make strong predictions about them.

Perhaps the most striking result is the similarity to the PDS of
Galactic XRBs.  
Besides having similar shapes, a direct scaling of mass with cutoff time 
scale (or luminosity) yields a black hole mass of 
$ \sim 10^6 - 10^7 M_\odot $ for NGC 3516.  
This may indicate that similar physical processes operate in both types of 
compact X-ray sources, spanning many orders of magnitude in luminosity.  If 
so, we could hope to apply what has been learned about XRBs to help 
understand (the more difficult to study) AGN.

Although the PDS shows a clear change in slope, the actual shape of
the low-frequency PDS is not well-determined.  
This can only be accomplished with data on even longer time scales, to 
sample even lower temporal frequencies than in this experiment.  
In fact, the 4.3~day sampling of NGC~3516 is continuing, and assuming that 
\xte\ continues to operate until at least April 2000, the PDS can be 
extended a factor of two lower in temporal frequency.  
Until these long time scales are sampled, it will not be possible to 
constrain the shape of the turnover.  
Indeed, characterizing the long time scale variability
properties and cutoff frequency is likely be the lasting legacy of
\xte\ for AGN, producing the first advance in our knowledge of PDS
since \exosat\ and showing that the most interesting action is on the
{\it longest}, not shortest accessible time scales.

\acknowledgments
As mentioned earlier, the near-perfect even sampling was achieved by
the heroic effort, well above the call of duty, of Evan Smith of the
\xte\ SOC.  
The entire \xte\ team put a great deal of effort into this project, without 
which it would not have been a success: 
Tess Jaffe and Gail Rorbach were instrumental in helping with the data 
reduction, Keith Jahoda, Dave Smith and Nick White helped with the 
background problems, Jean Swank scheduled the 1997 September-November 
observations that allowed smooth transition to the 4.3 day monitoring, and 
Alan Smale helped with a variety of issues.  
David Brillinger of the Berkeley Statistics Department also made a key 
contribution by explaining to us the statistics of combining data sampled 
on multiple time scales.
Simon Vaughan and Mark Dixon helped with the data analysis.
RE acknowledges financial support through NASA \xte\ grant NAG 5-7315 
and contract S-92507-Z, and KN acknowledges support through NASA grant 
NAG 5-7067 and the Universities Space Research Association.

\begin{figure}
\figurenum{1}
\epsscale{0.8}
\hspace{-1.5cm}
\plotone{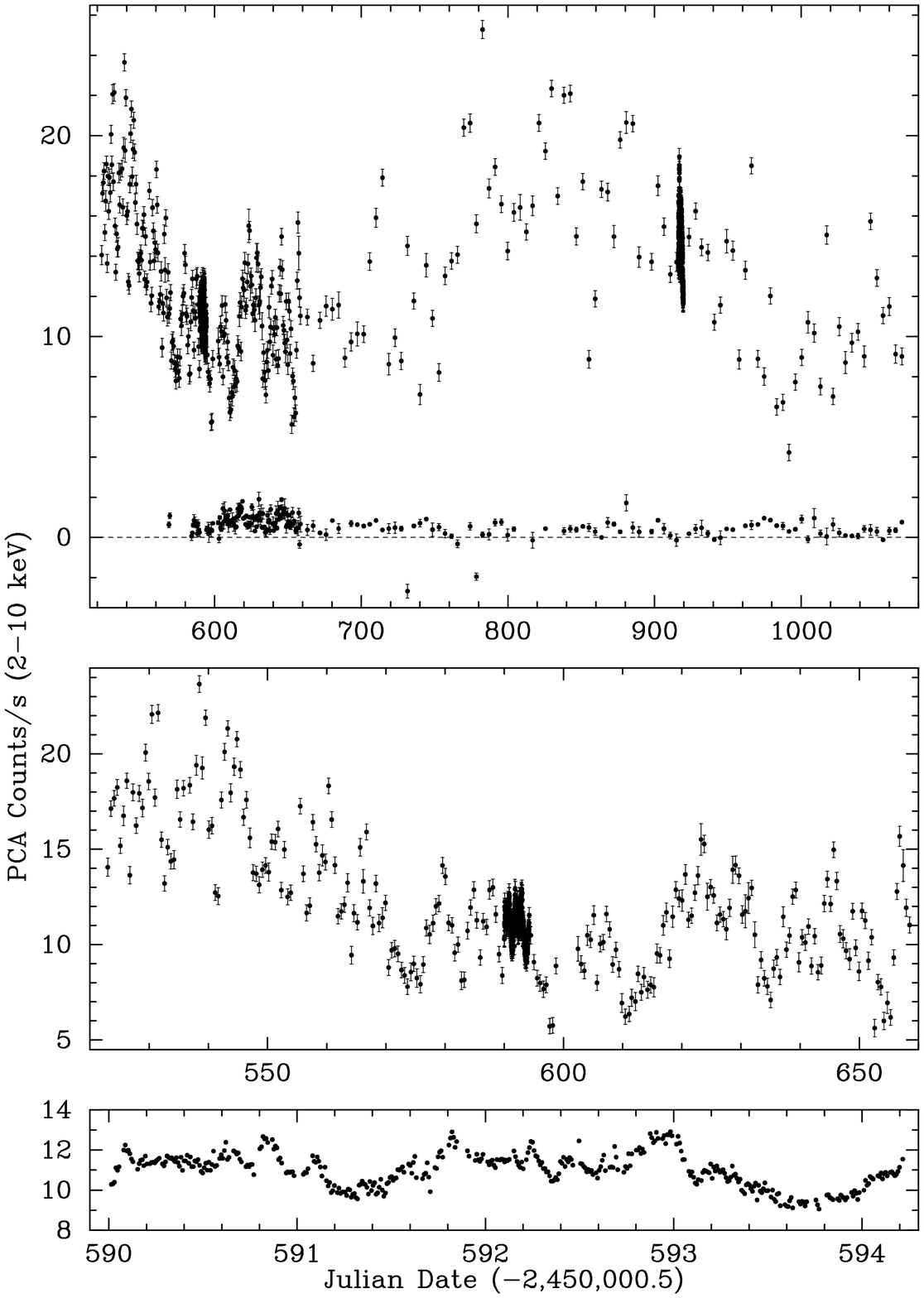}
\vspace{-2cm}
\caption{\xte\ light curve of NGC~3516.
The top panel covers the period 1997 March 16--1998 September 12.
Starting on 1997 July 31, the light curve was sampled every $\sim$4.3~days.
The data at the top is the source light curve while those at the bottom
(--1 to +2 c/s) are off-source (background pointings).
The dense sampling around MJD = 920 was obtained for a simultaneous {\it HST} 
campaign, which will be discussed in a future paper (Edelson \et 1998).
The middle panel covers the period 1997 March 16 to July 30, during
which time observations were made every $\sim$12.8~hr.
The bottom panel shows the data during 1997 May 22 00:01--May 26 05:37,
during which time the source was observed quasi-continuously, with
interruptions only due to SAA passage and occasional conflicting
time-critical observations, binned every 710~sec.
The abscissa is the background-subtracted PCA 2--10~keV L1 count
rate.
The error bars are a combination of statistical and systematic
uncertainties in the background subtraction.
(The latter generally dominate; see text for details.)
No error bars are shown for the short time scale data because they would make 
the figure too crowded.
However, they are generally the same as in the upper two panels.}
\label{fig1}
\end{figure}

\begin{figure}
\figurenum{2a}
\epsscale{1.0}
\vspace{-2cm}
\plotone{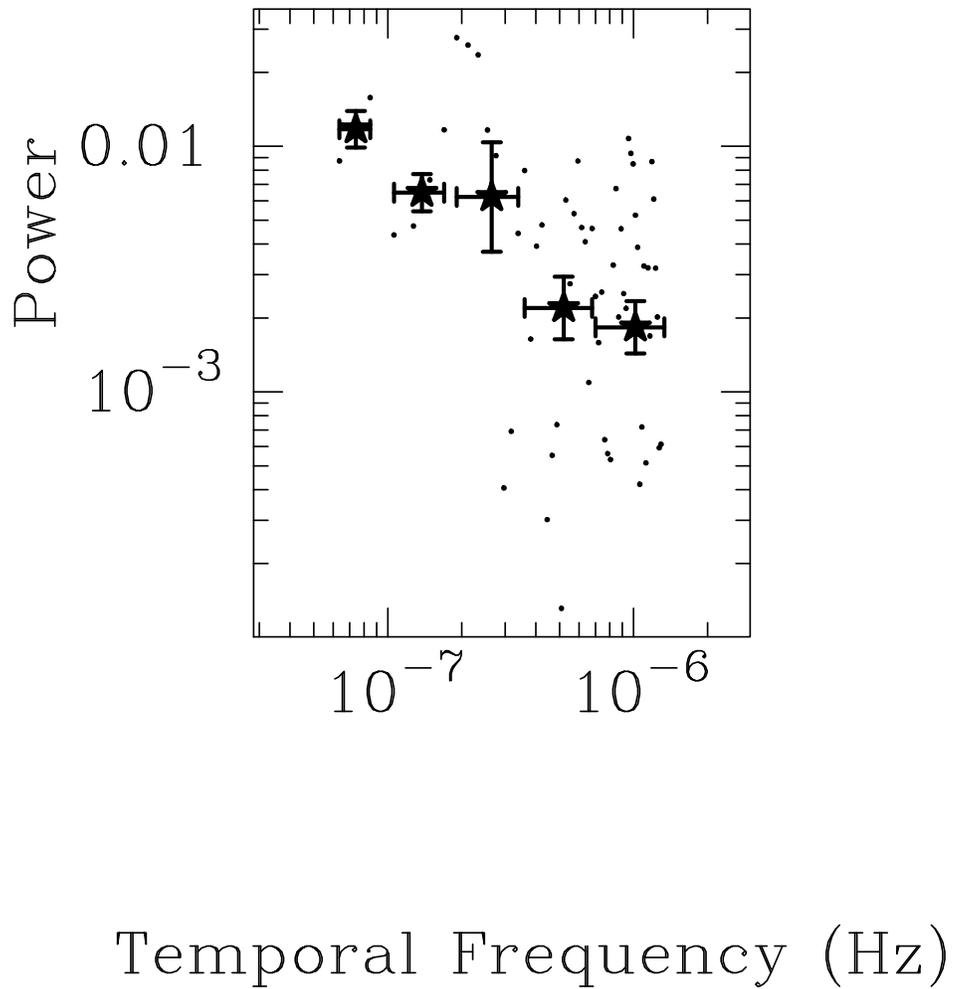}
\vspace{-2cm}
\caption{PDS derived from the long time scale data.
The first three points of the PDS have been suppressed as they
represent the long-term power that is not accurately determined.
The dots are the individual points in the PDS, while the large symbols
are results of logarithmic binning.
There is no evidence for periodicity, and the binned points were fitted
with a power-law model, as discussed in the text.
\label{fig2a}}
\end{figure}

\begin{figure}
\figurenum{2b}
\epsscale{1.0}
\vspace{-2cm}
\plotone{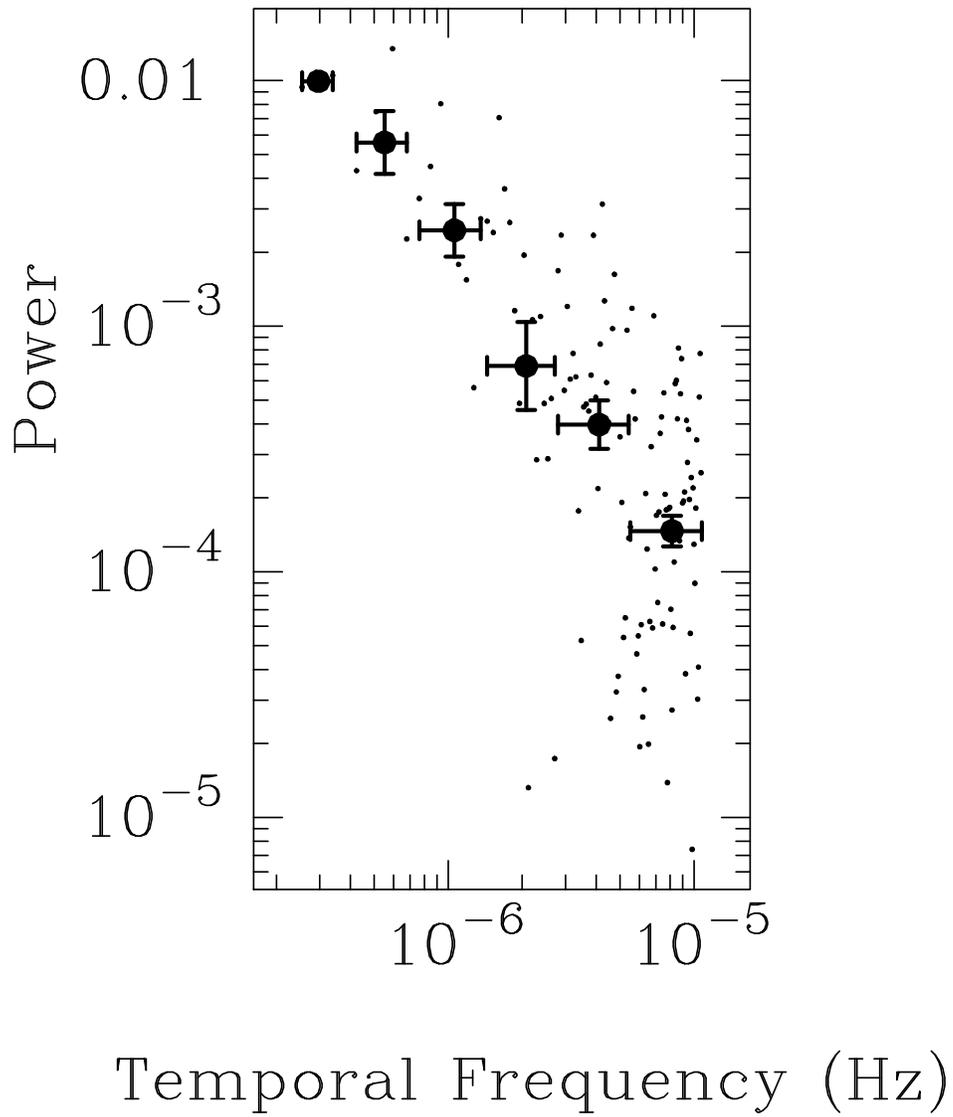}
\caption{Same as Figure~2a, except for the medium time scale data.
\label{fig2b}}
\end{figure}

\begin{figure}
\figurenum{2c}
\epsscale{1.0}
\vspace{-2cm}
\plotone{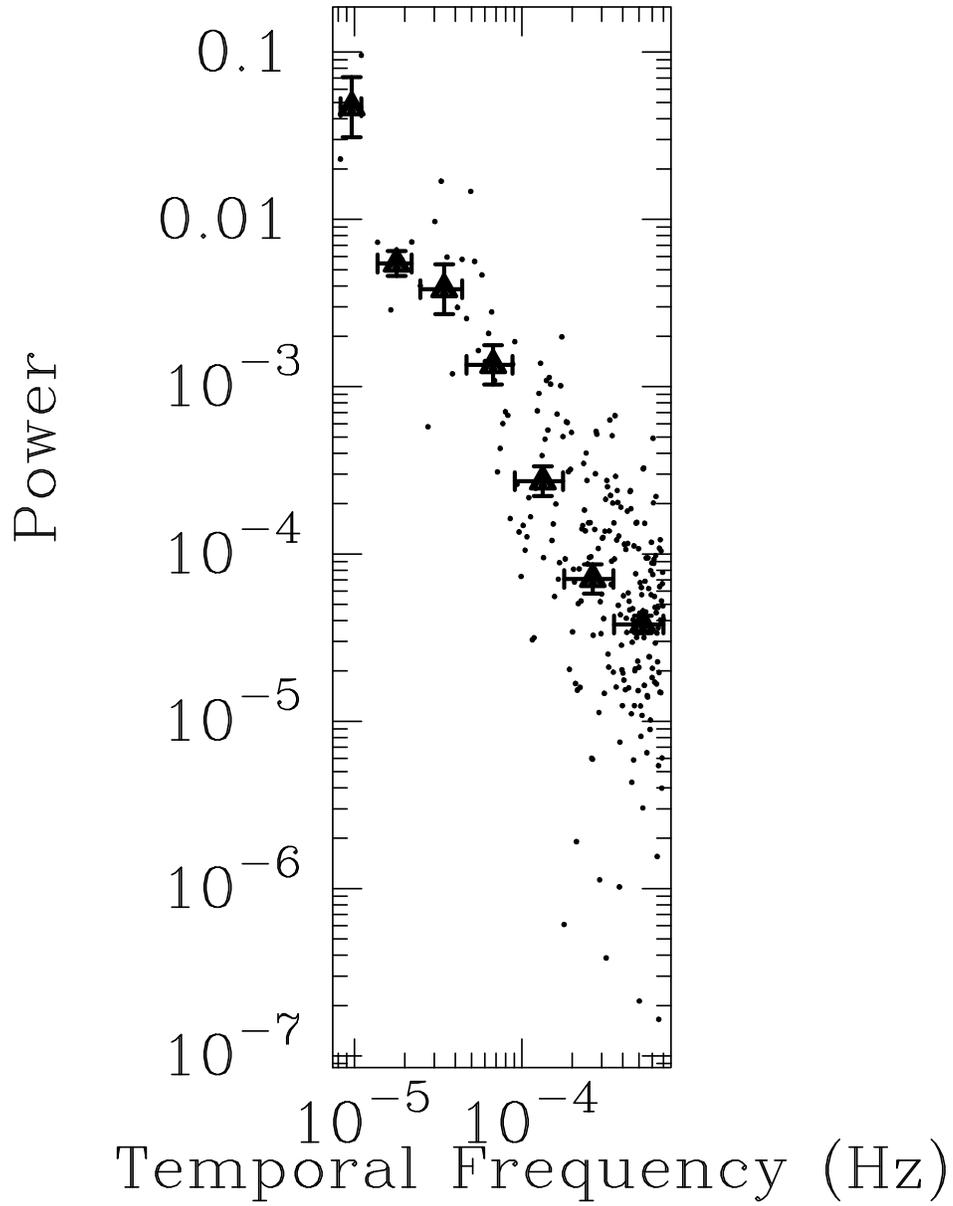}
\vspace{-2cm}
\caption{Same as Figure~2a, except for the short time scale data.
\label{fig2c}}
\end{figure}

\begin{figure}
\figurenum{3}
\epsscale{1.0}
\vspace{-2cm}
\plotone{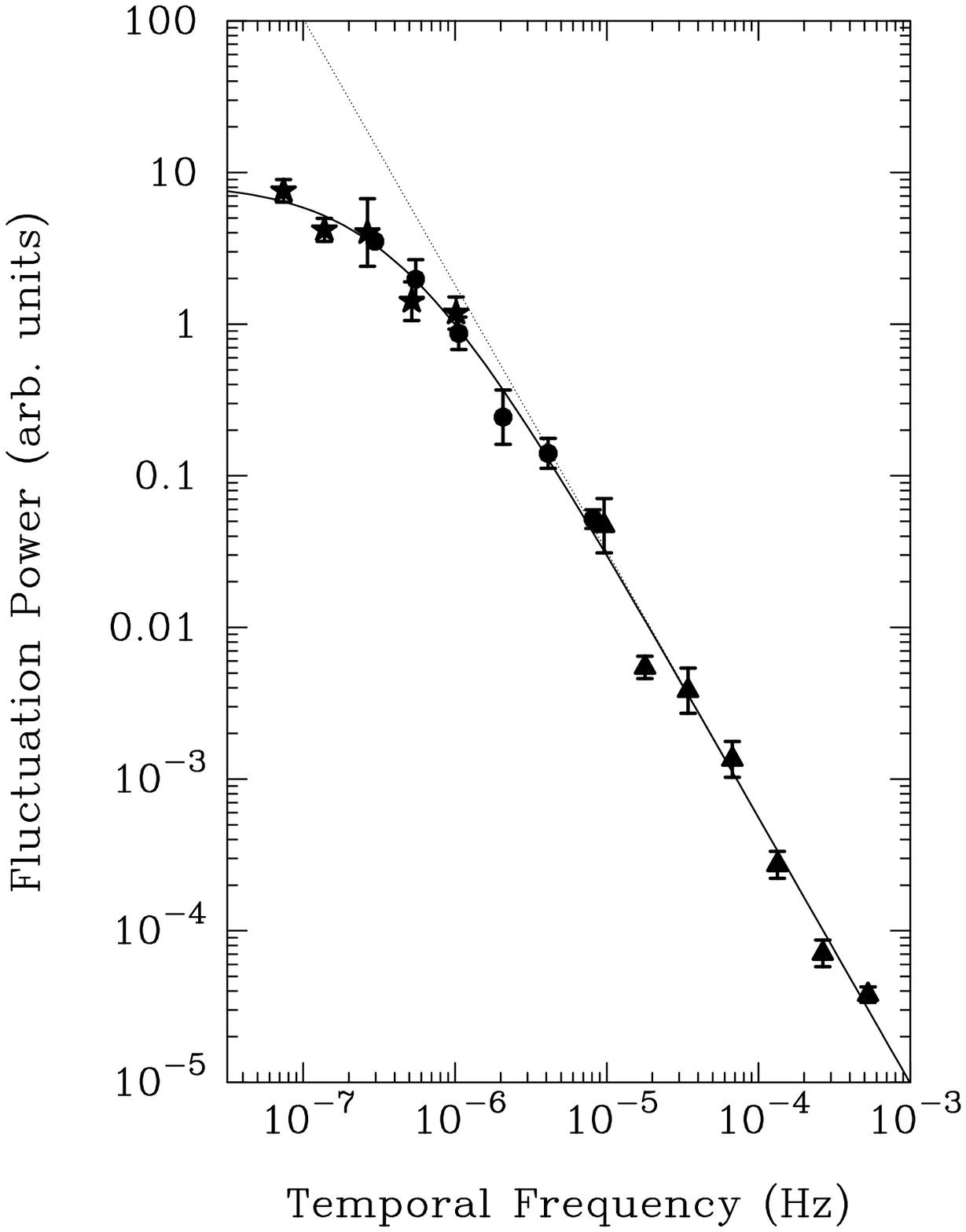}
\vspace{-2cm}
\caption{Broadband PDS of NGC~3516.
The individual PDS in Fig.~2 were combined, scaled and fitted 
Simultaneously, as described in the text.
The plot symbols follow the convention in Fig.~2.
The PDS is not consistent with a single power-law; instead it had to be 
fitted by a power-law that flattened to low frequencies in order to obtain 
a reasonable fit. 
This yielded a cutoff frequency of $ \sim 4 \times 10^{-7} $~Hz, 
corresponding to $\sim$1 month.
\label{fig3}}
\end{figure}


\begin{references}
\reference{B96} Bao, G.\ \& Abramowicz, M.\ 1996, ApJ, 465, 646
\reference{B90} Belloni, T.\ \& Hasinger, G.\ 1990, A\&A, 230, 103
\reference{B87} Brainerd, J.\ \& Lamb, F.\ 1987, ApJ, 317, L33
\reference{B88} Bussard, R.\ \et 1988, ApJ, 327, 284
\reference{D82} Deeter, J.\ \& Boynton, P.\ 1982, ApJ, 261, 337
\reference{E96} Edelson, R.\ \et 1996, ApJ, 470, 364
\reference{E96} Edelson, R.\ \et 1998, in preparation
\reference{G98} Ford, H.\ \et 1994, ApJ, 435, L27
\reference{G98} George, I., Turner, T.\ J., Netzer, H., Nandra, K., 
        Mushotzky, R., Yaqoob, T.\ 1998, ApJS, 114, 73 
\reference{G91} Ghosh, K.\ \& Soundararajaperumal, S.\ 1991, \apj, 383, 574
\reference{G93} Green, A., McHardy, I.\ \& Lehto, H.\ 1993, MNRAS, 265, 664
\reference{H91} Haardt, \& Maraschi, L.\ 1991, ApJ, 380, L51
\reference{H93} Haardt, \& Maraschi, L.\ 1993, ApJ, 413, 507
\reference{H95} Herrero, A., Kudritzki, R., Gabler, R., Vilchez, J.,
	Gabler, A.\ 1995, A\&A, 297, 556	
\reference{H97} Hua, X.-M., Kazanas, D., Titarchuk, L.\ 1997, ApJ, 482, L57
\reference{I98} Iwasawa, K., Fabian, A., Brandt, W.\ N., Kunieda, H.,
	Misaki, K., Reynolds, C.\ \& Rerashima, Y.\ 1998, MNRAS, 295L, 20
\reference{J96} Jahoda, K., Swank, J., Giles, A., Stark, M., 
        Strohmayer, T., Zhang, W., Morgan, E.H.\ 1996,
        EUV, X-ray and Gamma-ray Instrumentation for Space Astronomy 
        VII, O. H. W. Siegmund and M. A. Grummin, Eds., SPIE 2808, p. 59
\reference{K97} Kazanas, D., Hua, X.-M., Titarchuk, L.\ 1997, ApJ, 480, 735
\reference{K93} Kriss, G.A., \et 1993, ApJ, 467, 629
\reference{K93} Kolman, M. \et 1993, \apj, 403, 592
\reference{K98} Koratkar, A.\ \et 1998, BAAS, 192, 1607
\reference{K87} Kylafis, N. \& Klimis, G.\ 1987, ApJ, 323, 678
\reference{L93} Lawrence, A.\ \& Papadakis, I.\ 1993, \apj, 414, L85
\reference{M87} McHardy, I.\ in ``Two Topics in X-ray Astronomy," Eds.\ H.\ 
	Hunt \& B.\ Battrick, ESA SP-296 (ESA:Noordwijk), p.\ 1111
\reference{M92} Maraschi, L., Molendi, S.\ \& Stella, L.\ 1992,
        MNRAS, 225, 27
\reference{M98} Magorrian, J., \et\ 1998, AJ, 115, 2285
\reference{M9b} Mineshige, S., Ouchi, N., Nishimoti, H.\ 1994,
        PASJ, 46, 97
\reference{M9a} Mineshige, S., Takeuchi, M., Nishimoti, H.\ 1994,
        ApJ, 435, L125
\reference{M92} Miyamoto, S., Kitamoto, S., Iga, S., Negoro, H., Terada, 
	K.\ 1992, ApJ, 391, 21L
\reference{M95} Miyoshi, M., Moran, J., Herrnstein, J., Greenhill, L.,
        Nakai, N., Diamond, P., Inoue, M.\ 1995, Nat, 373, 127
\reference{N9a} Nandra, K., George, I., Mushotzky, R., Turner, T.\ J.,
        Yaqoob, T.\ 1997a, ApJ, 477, 602
\reference{N9b} Nandra, K., George, I., Mushotzky, R., Turner, T.\ J.,
        Yaqoob, T.\ 1997b, MNRAS, 282, L7
\reference{N98} Nandra, K., Clavel, J., Edelson, R., George, I., Malkan, M., 
Mushotzky, R., Peterson, B.\ \& Turner, T.\ J.\ 1998, ApJ, 505, 549
\reference{N94} Nandra, K.\ \& Pounds, K.\ 1994, 268, 405
\reference{P93} Papadakis, I.\ \& Lawrence, A.\ 1993, Nature, 361, 233
\reference{PL95} Papadakis, I.\ \& Lawrence, A.\ 1995, MNRAS, 272, 161
\reference{PM95} Papadakis, I.\ \& McHardy, I.\ 1995, MNRAS, 273, 923
\reference{P78} Press, W.\ 1978, Comments on Astrophysics, 7, 103
\reference{P92} Press, W.\ \et 1992, ``Numerical Recipes: The Art of
        Scientific Computing," 2nd Edition, (Cambridge University
        Press)
\reference{R84} Rees, M.\ 1984, ARA\&A, 22, 472
\reference{S43} Seyfert, K.\ 1943, \apj, 97, 28
\reference{S72} Shakura, N.\ \& Sunyaev, R.\ 1973, A\&A, 24, 337
\reference{S95} Stern, B., Poutanen, J., Svensson, R., Sikora, M., 
        Begelman, M.\ 1995, ApJ, 449, L13
\reference{P96} Tagliaferri, G.\ 1996, \apj, 465, 181
\reference{T95} Terlevich, R.\ \et 1995, MNRAS, 95, 272, 198
\reference{T94} Titarchuk, L., Mastichiadis, A.\ 1994, ApJ, 433, L33
\reference{T88} Treves, A., Maraschi, L., Abramowicz, M.\ 1988, PASP, 100, 
427
\reference{V95} van der Klis, M. 1995, in X-Ray Binaries, ed. W. Lewin, 
        J. van Paradijs, \& van E. van den Heuvel 
        (Cambridge: Cambridge Univ. Press), 252
\reference{V98} Van Paradijs, J.\ 1998, in ``The Many Faces of Neutron 
	Stars," Eds.\ R.\ Buccheri \et (Kulwar), in press, astro-ph/9802177
\reference{V87} Voit, M., Shull, J.\ M.\ \& Begelman, M.\ 1987, 
        \apj, 316, 573
\reference{W93} Wanders, I., \et 1993, A\&A, 269, 39
\reference{Z94} Zdziarski, A., Fabian, A., Nandra, K., Celotti, A., Rees, 
	M., Done, C., Coppi, P., Madejski, G.\ 1994, MNRAS, 269, L55
\end{references}
\end{document}